%% file: thaller-fl-psm-workshop.tex
\documentclass[10pt, conference]{IEEEtran}
\IEEEoverridecommandlockouts

\usepackage{setspace}
\usepackage[utf8x]{inputenc}
\usepackage[T1]{fontenc}
\usepackage[english]{babel}
\usepackage{microtype}
\usepackage{amsmath,amssymb,amsfonts, bm, amsthm}
\usepackage{algorithmic}
\usepackage{graphicx, tikz}
\usepackage{caption, subcaption}
\usepackage{booktabs, threeparttable, multirow}
\usepackage{stfloats}
\usepackage[inline]{enumitem}
\usepackage{url}
\usepackage{xcolor}
\usepackage{hyperref}
\usepackage{pifont}
\usepackage[framemethod=tikz]{mdframed}
\usepackage{siunitx}
\usepackage{balance}

\graphicspath{{images/}, {scripts/}}
\DeclareGraphicsExtensions{.pdf,.jpeg,.png}

\definecolor{dark-red}{rgb}{.6, .15, .15}
\definecolor{dark-blue}{rgb}{.15, .15, .55}
\definecolor{accent1}{HTML}{24B9FC}
\definecolor{accent2}{HTML}{9ECD67}
\definecolor{accent3}{HTML}{FFA929}
\definecolor{light-bg}{HTML}{B2B2B2}

\newcommand{\code}[1]{\texttt{#1}}
\newcommand{\ra}[1]{\renewcommand{\arraystretch}{#1}}

\hypersetup{
	colorlinks,
  linkcolor={dark-red},
	citecolor={dark-blue},
  urlcolor={black}
}

\newif\ifunblind
\unblindtrue

\begin{document}

\title{Towards Fault Localization \\ via Probabilistic Software Modeling}

\ifunblind
\author{%
	\IEEEauthorblockN{Hannes Thaller, Lukas Linsbauer, Alexander Egyed}
	\IEEEauthorblockA{Institute for Software Systems Engineering\\
		Johannes Kepler University Linz, Austria\\
		\{hannes.thaller, lukas.linsbauer, alexander.egyed\}@jku.at} \and
	\IEEEauthorblockN{Stefan Fischer}
	\IEEEauthorblockA{Software Competence Center Hagenberg GmbH\\
		Austria\\
		stefan.fischer@scch.at}
}
\else
\author{\IEEEauthorblockN{Author}
	\IEEEauthorblockA{Institute\\
		Affiliation, Country\\
		email@address.com}
}
\fi

\maketitle

\begin{abstract}
Software testing helps developers to identify bugs.
However, awareness of bugs is only the first step.
Finding and correcting the faulty program components is equally hard and essential for high-quality software.
Fault localization automatically pinpoints the location of an existing bug in a program.
It is a hard problem, and existing methods are not yet precise enough for widespread industrial adoption.
We propose fault localization via Probabilistic Software Modeling (PSM).
PSM analyzes the structure and behavior of a program and synthesizes a network of Probabilistic Models (PMs).
Each PM models a method with its inputs and outputs and is capable of evaluating the likelihood of runtime data.
We use this likelihood evaluation to find fault locations and their impact on dependent code elements.
Results indicate that PSM is a robust framework for accurate fault localization.
\end{abstract}


\IEEEpeerreviewmaketitle

\input{main}

\ifunblind
\section*{Acknowledgments}
The research reported in this paper has been supported by the Austrian ministries BMVIT and BMDW, and the Province of Upper Austria in terms of the COMET - Competence Centers for Excellent Technologies Programme managed by FFG.
\fi

\bibliographystyle{IEEEtran}
\bibliography{references}

\end{document}

%% file: main.tex





\section{Introduction}
Modern software development aims to design and control the quality of software.
Testing techniques, such as unit, integration, or system testing, and their automation via continuous integration, provide a feasible and generally applicable approach for software quality assurance.
Software testing aims to find faults in a program.
However, tests can not localize the faults within a program's source code.
This is no issue for unit testing since the tests are small enough (typically methods).
However, fault localization for integration and system tests can become a time-consuming task.

Fault Localization (FL) is the task of automatically finding faults in a program such that a developer or an automated process can repair them.
Finding a fault, i.e., the real cause of an error, is a hard problem.
Not only is it difficult to distinguish a symptom (cascading error) from a cause (actual fault), but also multiple faults can work in conjunction, complicating the localization process.
The state-of-the-art FL techniques like Spectrum-based Fault Localization (SBFL)~\cite{Jones2002, Wong2014} traditionally rank statements by their likelihood of containing a fault.
This leads to localization weaknesses for complex faults that span multiple lines~\cite{Wong2016} (76\% of faults) or that are caused by the omission of statements (30\% of faults)~\cite{Pearson2017, Parnin2011}.

We propose Fault Localization via Probabilistic Software Modeling (FL-PSM).
PSM~\cite{Thaller2019e} builds a network of Probabilistic Models (PMs) of the executables (e.g., methods in Java) in a program.
We use the PMs built by PSM to locate the most likely fault location.
FL-PSM is a dynamic approach using either the test-suite (such as test-based FL techniques) or the actual execution of a program to fit each PM.
PSM uses runtime data to construct behavioral datasets with which it fits the PMs.
Then, runtime data from another program version, or failing tests, are used to find the most likely fault location.

\section{Running Example}
\begin{figure*}[ht]
    \centering
    \includegraphics[width=\textwidth]{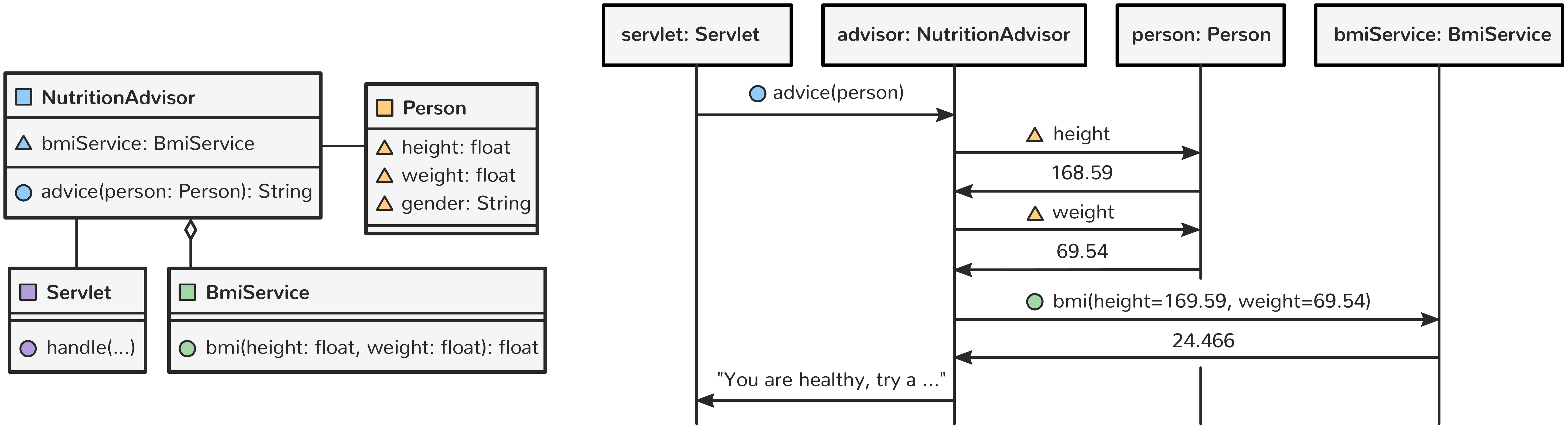}
    \caption{Class Diagram (left) and Sequence Diagram (right) of the Nutrition Advisor~\cite{Thaller2019e}.}
    \label{fig:nutrition advisor}
\end{figure*}
We use as an illustrative example the \emph{Nutrition Advisor} that takes a person's anthropometric measurements (e.g., height and weight) and returns a piece of textual advice based on the Body Mass Index (BMI).
The class diagram in Figure~\ref{fig:nutrition advisor} (left) shows the four classes of the Nutrition Advisor.
The sequence diagram in Figure \ref{fig:nutrition advisor} (right) shows a possible runtime trace of a request handled by the program.
The \code{Servlet} handles requests (\code{handle()}) and initializes a \code{Person} object.
This \code{Person} object is received by the \code{NutritionAdvisor.advice}-method that extract the person's height (168.59) and weight (69.54).
Both values are the parameters for the \code{BmiService.bmi} call that returns the BMI (24.466) with which a textual advice is returned ("You are healthy, \ldots").

\section{Background}
Probabilistic Software Modeling~\cite{Thaller2019e} describes a methodology for transforming a program into a network of probabilistic models.
It extracts a program's structure represented by properties, executables, and types (fields, methods, and classes in Java) along with their call dependencies to build a network of probabilistic models.
Every node in the network is a PM that represents an executable.
Each PM in the network is optimized towards a program execution.
These execution traces are extracted from the system in its production environment or triggered via tests.

Each code element PM represents an executable (e.g., a Java method) in the program.
Inputs are parameters, property reads, invocation return values, while outputs are the method return value, property writes, and invocation parameters.
The distinction between inputs and outputs exists only on a logical level for the program.
However, the models themselves are multivariate density estimators (unsupervised models) with no notion of input and output (joint model of all variables).
Each model can generate new observations that are similar to the initially trained data, e.g., to generate likely or rare (but plausible) test data.
Furthermore, each model can evaluate the likelihood of a given observation (e.g., to evaluate the adequacy of given test data).
This evaluation is relative to the runtime trace that was used to fit the model, e.g., a model based on production runtime will evaluate observations differently than a model based on tests.

PMs in this work are Non-Volume Preserving Transformations (NVPs)~\cite{Dinh2016, Papamakarios2019a}, which are general and expressive flow-based density estimators.
Each NVP is built via neural networks that learn a function that maps latent random variables (e.g., Gaussian variables) to the data (runtime events).
Evaluating the likelihood with NVPs is done by transforming the runtime events into the known Gaussian latent-space and computing the Gaussian likelihood of the transformed events. 
More details on PSM and NVPs are given by our previous work \cite{Thaller2019e} and Dinh~\cite{Dinh2016, Papamakarios2019a}.

\section{Approach}\label{sec:approach}
FL-PSM is built upon PSM. The fault localization is based on the likelihood evaluation of these models.
Given is a null-model $M^{null}$ of an executable and either an alt-dataset $D^{alt}$ of runtime events or an alt-model $M^{alt}$ with which a dataset is generated.
FL-PSM localizes faults by computing the mean log-likelihood of $D^{alt}$ on $M^{null}$ and comparing it to a critical value.
More specifically,
\begin{equation}
    LL_{D^{alt}} = \dfrac{1}{N} \sum_i^N p_{M^{null}}\left(D^{alt}_i\right)
\end{equation}
computes the average log-likelihood where $N$ is the number of data points in $D$.
Finally, 
\begin{equation}
    LL_{D^{alt}} - LL_{D^{null}} < c
\end{equation}
evaluates whether there exists a significant difference between model and data.
$LL_{D^{null}}$ is the log-likelihood of $M^{null}$ to itself and captures the inherent bias.
The critical value $c$ controls for Type-1 errors (false-positives) similar to other significance tests, e.g., $\log(0.001)$ indicates that 1 out of 1000 events is falsely considered to be significantly different from the model.

\section{Preliminary Study}
This preliminary study shows how FL-PSM finds possible fault locations.
Given is the Nutrition Advisor to which 3000 requests are made based on data from the NHANES~\cite{Cdc2013} dataset.
The resulting model is the null-model $M_{null}$.
Then we seeded two errors in the Nutrition Advisor and collected the alt-datasets $D1_{alt}$ and $D2_{alt}$.
The first error simulates a regression (between versions) caused by a typo in the \texttt{Person} constructor that assigns \texttt{-weight} instead of \texttt{weight} to the field.
The second error simulates an integration fault (within version) caused by the miscommunication between teams using different measures.
Team A that also built the \emph{null} Nutrition Advisor, computes the BMI in meters while Team B that revises the implementation computes the BMI in inches.

We used the computation from Section~\ref{sec:approach} with a critical value (i.e., false-positive rate) of $c = \log(0.001) = -3$.
This means log-likelihoods below $-3$ are significantly diverging from the model.

\subsection{Regression Fault}
\begin{figure*}[ht]
    \centering
    \includegraphics[width=.7\textwidth]{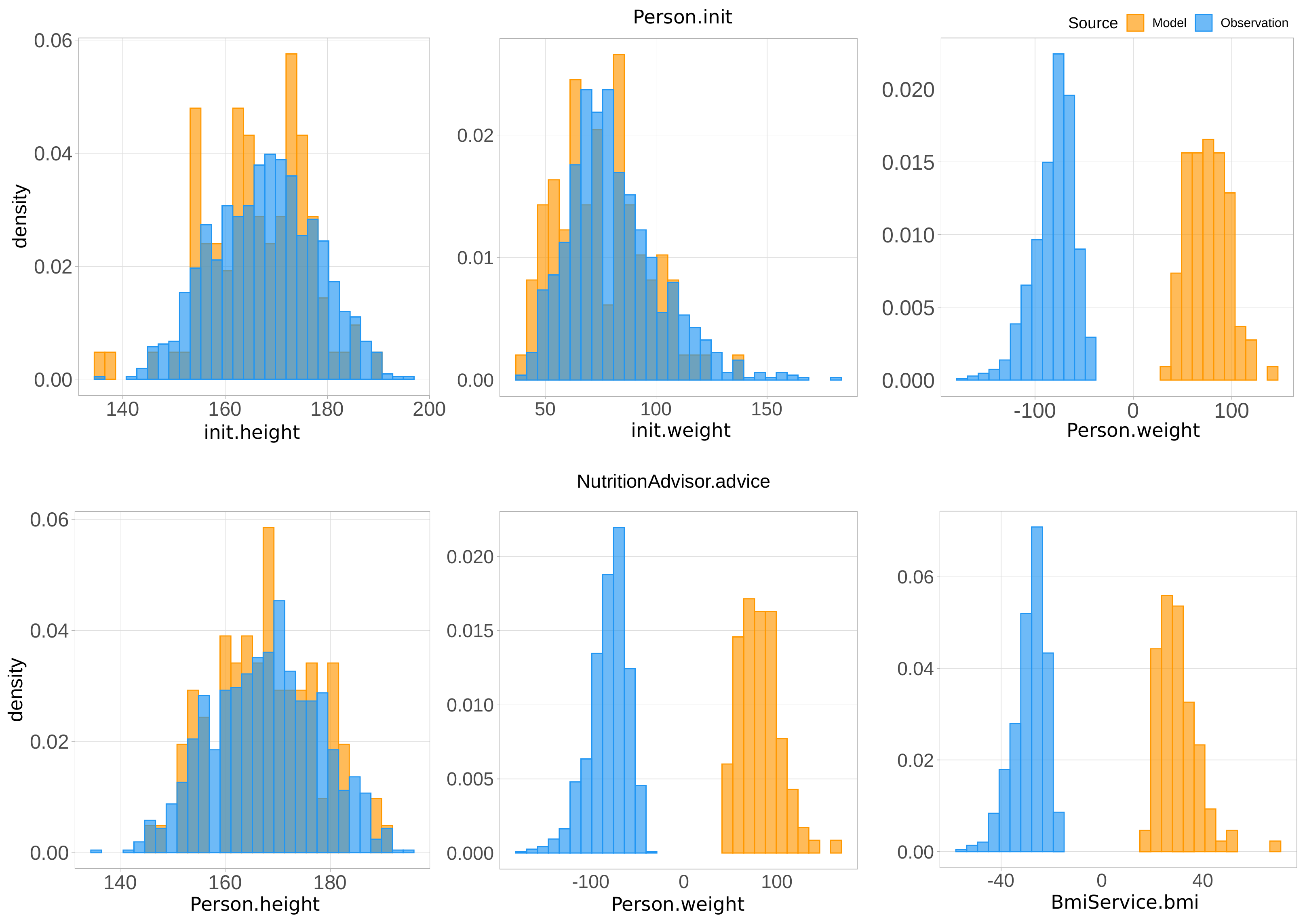}
    \caption{
    A subset of elements in the regression fault setting.
    For example, the first row shows \emph{Person.init} parameters and property writes between the original version and a regressed version of the same component.
    }
    \label{fig:regression}
\end{figure*}
Figure~\ref{fig:regression} shows the runtime behavior of a subset of code elements of the \emph{Person.init} and \emph{NutritionAdvisor.advice} models.
Table~\ref{tab:regression} lists the likelihood and significance of these elements along with the multivariate model likelihood that considers all elements at once.
The visualization of the code elements allows developers to see that there is a significant difference between the model and the observations.
The constructor parameter \emph{init.weight} is aligned with the model while the property writes to \emph{Person.weight} are clearly different.
This difference is also significant as Table~\ref{tab:regression} shows (rows 1 and 4).
Other elements in the same model are insignificantly different as both the visualization and the table show.

The difference propagates to the depending \emph{NutritionAdvisor.advice} method that reads the \emph{Person.weight} property (rows 5 and 7).
Also the invocation of the \emph{BmiService.bmi} indicates this significant divergence (row 8).

\input{tab_regression.tex}

\subsection{Integration Fault}
\begin{figure*}[ht]
    \centering
    \includegraphics[width=.7\textwidth]{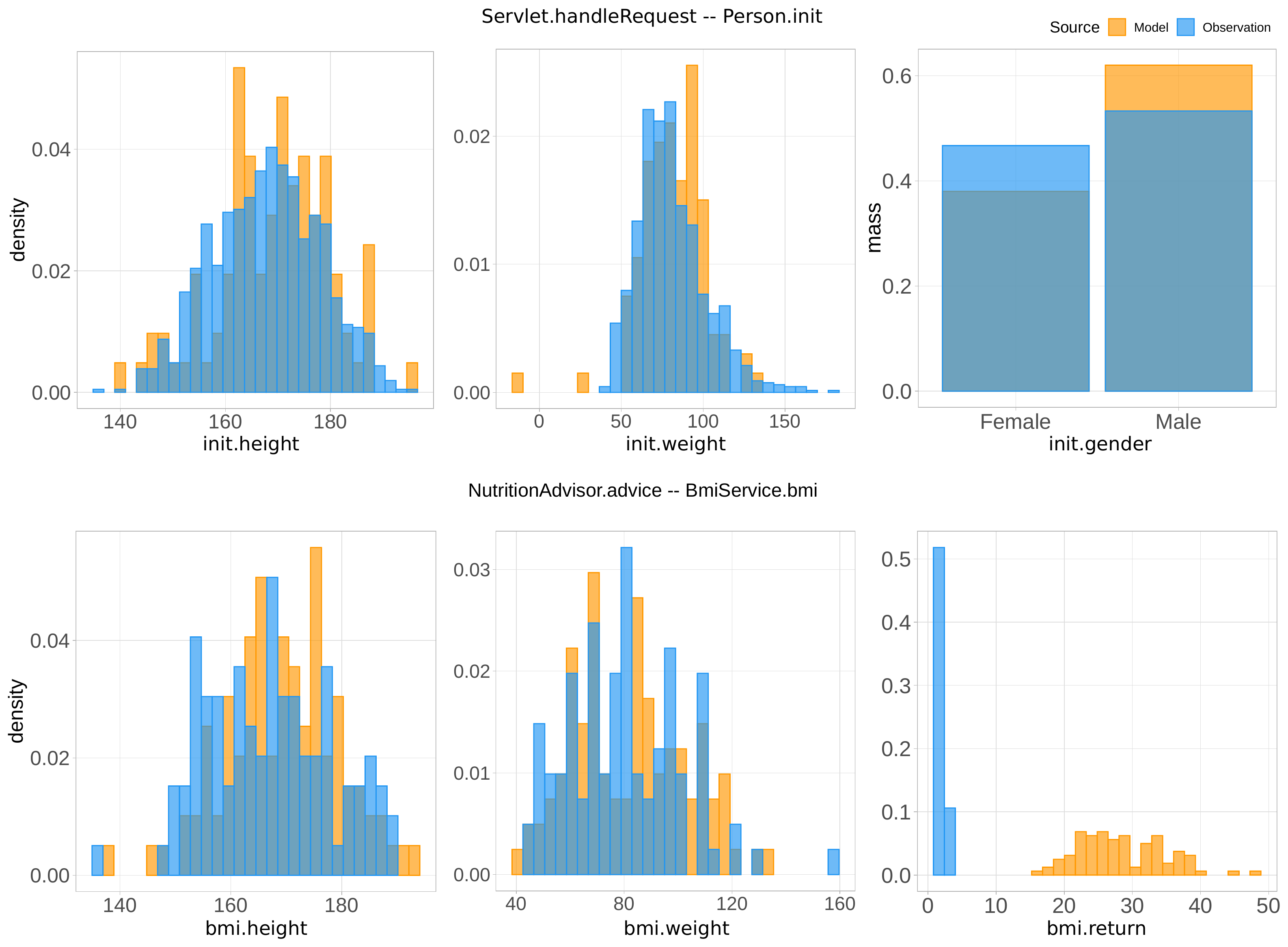}
    \caption{
    A subset of elements in the integration fault setting.
    For example, the first row shows \emph{Person.init} parameter values caused by the invocation from \emph{Servlet.handleRequest}.
    }
    \label{fig:integration}
\end{figure*}
Figure \ref{fig:integration} shows the runtime behavior of a subset of code elements of the \emph{Servlet.handleRequest} calling \emph{Person.init}.
In this case, \emph{Servlet.handleRequest} model evaluates parameters or return values of \emph{Person.init}.
The visualization shows no significant difference in the integration between the \emph{Servlet} and \emph{Person}.
This insignificance is also given in Table~\ref{tab:integration}.
The integration between \emph{NutritionAdvisor.advice} and \emph{BmiService.bmi}, with the first being the model, shows a difference in the return value of \emph{BmiService.bmi}.
Again, this difference is also reflected in Table~\ref{tab:integration} (rows 5 and 8).

\input{tab_integration.tex}

\section{Discussion}
The preliminary study showed how FL-PSM localizes faults.
This localization is automated via likelihood-based significance tests that allow for statistical control of the false-positive rate.
The other important aspect is the visualization of the faults (Figures \ref{fig:regression} and \ref{fig:integration}) and its impact on dependent elements.
This allows for precise analysis of the error chain and its influence across the program.

FL-PSM can only be applied if there is at least a version of the program.
This is not an issue from an industrial point of view since FL-PSM can be used after a few development sprints.
Another consideration is that FL-PSM localizes behavioral changes, including intended changes.
These intended changes can be filtered by incorporating source code change information in the localization process.
In addition, the visualization capabilities of FL-PSM allow for quick manual inspections in cases of doubt.

In summary, the results and usability of FL-PSM are promising.
Nevertheless, there are still open questions concerning multiple fault sources and their clear separation.

\section{Related Work}
Most fault localization techniques are slice, spectrum, statistics, model, or machine learning-based~\cite{Jones2002, Wong2016}.

The most similar technique to FL-PSM is Spectrum-Based Fault Localization (SBFL)~\cite{Jones2002}.
SBFL techniques observe passing and failing executions and perform statistical inference on the results.
The result is a ranked list of statements, along with their likelihood of being the fault location.
While similar, FL-PSM works slightly differently in terms of the abstraction level and source model.
PSM abstracts statements and only considers properties, executables, and types along with their call dependencies.
In contrast, SBFL techniques predominately work on the statement level.
This might seem like a drawback at first.
However, Parnin and Orso~\cite{Parnin2011} identified that the detail of the results in combination with high false-positive rates are one of the main issues of the low industrial adoption of SBFL.
PSM improves on these issues by providing control of the false-positive rate and its level of abstraction (executables).

\section{Conclusion and Future Work}
We presented Fault Localization via Probabilistic Software Modeling (FL-PSM).
FL-PSM builds upon PSM and uses statistical inference to find possible fault locations in a program.
The localization is based on evaluating the likelihood of runtime events under the model.
We have shown how FL-PSM localizes and visualizes faults.
In addition, we discussed the difference between FL-PSM and its close relative SBFL.

Future work will focus on a full evaluation of the approach with multiple complex subsystems.
Furthermore, we want to conduct a user study for its practicality and applicability.

In conclusion, FL-PSM is a promising new FL approach built upon PSM that provides a general framework for probabilistic analysis of software programs.

%% file: tab_regression.tex
\begin{table}[!t]
    \ra{1.2}
    \centering
    \footnotesize
        \caption{
    Likelihood values of a subset of elements in the regression fault setting.
    }
    \label{tab:regression}
    \begin{threeparttable}[b]
    \begin{tabular}{@{}c l r r r r @{}}
        \toprule
        & Model & Element & Cardinality &  LL & Sig \\
        \midrule
        1 & Person.init & init  & multivariate &  -6787  & \ding{52} \\
        2 & ~Person.init & init.height & univariate &  -1.74 & \ding{54} \\
        3 & ~Person.init & init.weight & univariate & -2.17 & \ding{54} \\
        4 & ~Person.init & Person.height & univariate & -49.49 & \ding{52} \\
        5 & NutritionAdvisor.advice & advice & multivariate & -4281 & \ding{52} \\
        6 & ~NutritionAdvisor.advice & Person.height & univariate & -0.96 & \ding{54} \\
        7 & ~NutritionAdvisor.advice & Person.weight & univariate & -97.23 & \ding{52} \\
        8 & ~NutritionAdvisor.advice & BmiService.bmi & univariate & -82.67 & \ding{52} \\
        \midrule
        \bottomrule
    \end{tabular}
    \end{threeparttable}
\end{table}

%% file: tab_integration.tex
\begin{table}[!t]
    \ra{1.2}
    \centering
    \footnotesize
        \caption{
    Likelihood values of a subset of element in the integration fault setting.
    }
    \label{tab:integration}
    \begin{threeparttable}[b]
    \begin{tabular}{@{}c l r r r r@{}}
        \toprule
        & Model  & Element & Cardinality &  LL & Sig. \\
        \midrule
        1 & Servlet.handle & Person.init  & multivariate &  0  & \ding{54} \\
        2 & ~Servlet.handle & init.height & univariate &  -1.46 & \ding{54} \\
        3 & ~Servlet.handle & init.weight & univariate & -2.33 & \ding{54} \\
        4 & ~Servlet.handle & init.gender & univariate & -1.95 & \ding{54} \\
        5 & NutritionAdvisor.advice & BmiService.bmi & multivariate & -6373 & \ding{52} \\
        6 & ~NutritionAdvisor.advice & bmi.height & univariate & -0.95 & \ding{54} \\
        7 & ~NutritionAdvisor.advice & bmi.weight & univariate & -0.50 & \ding{54} \\
        8 & ~NutritionAdvisor.advice & bmi.return & univariate & -13.22 & \ding{52} \\
        \midrule
        \bottomrule
    \end{tabular}
    \end{threeparttable}
\end{table}